\documentclass[conference]{IEEEtran}
\IEEEoverridecommandlockouts
\usepackage{cite}
\usepackage{amsmath,amssymb,amsfonts}
\usepackage{algorithmic}
\usepackage{graphicx}
\usepackage{textcomp}
\usepackage{xcolor}
\usepackage{booktabs, tabularx, multirow}
\usepackage[braket, qm]{qcircuit}
\usepackage{subcaption}
\usepackage{balance}
\def\BibTeX{{\rm B\kern-.05em{\sc i\kern-.025em b}\kern-.08em
    T\kern-.1667em\lower.7ex\hbox{E}\kern-.125emX}}
\begin{document}

\bstctlcite{IEEEexample:BSTcontrol}

% \title{Mutation Testing of Quantum Neural Networks\\
% Directed Mutation Testing of Quantum Machine Learning Models}
\title{Efficient Mutation Testing of Quantum Machine Learning Models}

\author{\IEEEauthorblockN{Emma Andrews and Prabhat Mishra}
\IEEEauthorblockA{
\textit{University of Florida}\\
Gainesville, Florida, USA
% \{e.andrews, prabhat\}@ufl.edu
}
}

\maketitle

\begin{abstract}
Quantum machine learning integrates the strengths of quantum computing and machine learning, enabling models to learn complex features using fewer parameters than their classical counterparts. Due to the increasing complexity of quantum machine learning models, it is necessary to verify that the implementation of these models satisfy the design specification and be free of bugs and faults. Mutation testing is a promising avenue to identify faulty quantum circuits that do not meet design specifications or contain defects by intentionally inserting faults into the quantum circuit. It is necessary to define mutation operations to inject faults into quantum circuits to ensure that a test suite is robust enough to evaluate an implementation against its  design specification. In this paper, we extend mutation testing to quantum machine learning applications, primarily quantum neural network models. Specifically, this paper makes two important contributions. We define new mutation operations for efficient fault insertion compared to state-of-the-art approaches. We also present a directed mutation generation technique to reduce redundant mutant circuits. Extensive experimental evaluation demonstrates that our approach generates a more diverse and representative set of mutants, effectively addressing faults that traditional techniques fail to expose.
\end{abstract}

% \begin{itemize}
% \item Existing solutions cover generic quantum circuits, and loses the opportunity for application-specific customizations. For example, it considers all possible basis states as inputs that leas to ... exponential. In contrast, we take advantage of the fact that QNN inputs are ... that drastically reduces the input complexity.
% \item Existing solutions need to consider all possible gates as potential mutation (any gate to any gate). In contrast, for QNN, the potential set of gates are limited (e.g., Rx, Ry, Rz). Moreover, Rx and Ry only changes amplitude, while Ry changes the phase.
% \end{itemize}


\section{Introduction}
% qc popularity, rising need for testing to ensure design specifications being met
% mutation testing a good approach, allows for creation of many faulty qcs or mutants
% these mutants can be used to evaluate the quality of a test suite in detecting faults in a design
% we focus on mutation operations in this work to generate a wide coverage of potential faults and to reduce the number of redudant mutants
% redundant mutants are those that produce the same faulty behavior, thus requiring only one of these faulty circuits to be executed to evaluate the test suite on
% reducing the number of redundant mutants reduces the time required to carry out the entire mutation testing process, requiring less simulation and mutation generation time

% prior state of the art has defined several basic mutation operations for quantum computing
% including adding, removing, or changing gates
% while these work well for general quantum circuits and do cover large portions of generalized functionality common to all quantum circuits, there are many applications and fields that define their quantum circuits in a specific manner where these mutation operations are not sufficient to produce all possible faulty behavior, thus providing an incomplete evaluation
Quantum computing is promising due to its ability to execute a specific classes of computations significantly faster than classical computing. Quantum machine learning (QML) is one of the application areas that can see the best use case of quantum computing over classical computing~\cite{schuld2015introduction}. QML has the potential to learn features of a dataset and how to perform a specific task with less computational resources compared to classical machine learning. With the increasing utilization of QML across application domains, it is necessary to test the resulting quantum circuits to ensure that they meet the design specification and are free of potential bugs.

\begin{figure}
    \centering
    \includegraphics[width=\linewidth]{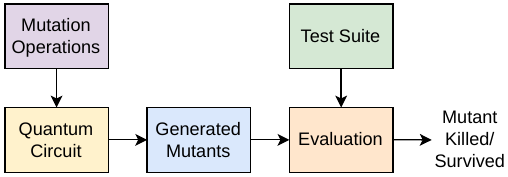}
    \caption{An overview of mutation testing.}
    \label{fig:mutation-testing}
    \vspace{-0.2in}
\end{figure}

While there are several strategies for testing quantum circuits, mutation testing is one of the most promising software testing techniques. Mutation testing takes a program and intentionally inserts faults into it, producing mutants or faulty programs~\cite{jia2011analysis}. The test suite should be able to detect all of these faulty programs since they will deviate from the design specification. In case of quantum computing (Figure~\ref{fig:mutation-testing}), mutation testing is performed specifically on the quantum circuits and often using test oracles as the test suites~\cite{mendiluze2021muskit, fortunato2022mutation, mendiluzeusandizaga2025quantum}.

Mutants are created through mutation operations, where an expected behavior of the quantum circuit is altered. For example, this may include mutation operations such as adding a new gate into the original quantum circuit, removing a gate from the original quantum circuit, or replacing a gate with a different functionality. In other words, once we apply a mutation operation, it produces a faulty quantum circuit, known as `\textit{mutant}'. Exhaustive generation of mutants can provide a complete set of faulty circuits using these mutations and covers the faulty scenarios with each mutation operation. Unfortunately, exhaustive generation can lead to exponential number of mutants and prohibitive validation time. 

There are two important considerations in mutation testing. A major challenge in mutation testing is how to effectively generate a small set of mutants that can cover a wide variety of potential faulty scenarios. Moreover, we need to account for the time required to test the generated mutants against the test suite. The test suite validation cost could be prohibitive if there are too many mutants, since validation of the generated mutants against the test suite involves running the quantum circuits on real hardware or on simulators. A promising avenue for cost reduction is to avoid generation of redundant mutants since they will  exhibit the same faulty behavior. 

State-of-the-art mutation testing for quantum circuits performs well at generating mutated circuits with faulty behaviors~\cite{mendiluze2021muskit, fortunato2022mutation, mendiluzeusandizaga2025quantum}. These mutation operations target normal features of quantum circuits, such as addition or deletion of gates. State-of-the-art mutation operations are insufficient in covering all potential faulty behaviors, even from a redundant mutant perspective, when the quantum circuits have advanced logic or application-specific structures. For example, quantum circuits for representing the QML models feature parameterized gates as well as specific structuring of the gates. 

New mutation operations are required to effectively represent all possible faulty behaviors in a QML model that cannot be captured by state-of-the-art mutation operations. Specifically, QML models contain two sections, a feature map to embed data into the model to learn from, and the ansatz, which performs the computations of the model. These two sections are structured in a specific manner unique to QML circuits, such as the ansatz featuring layer computations, that must be accounted for when performing mutation operations to generate mutants covering all possible faulty behaviors. To reduce the computation overhead, we also need to ensure that redundant mutants are not generated.

To address the above challenges for mutation testing of QML models, we develop new mutation operations to specifically target quantum neural network (QNN) models, including both the feature mapping and the ansatz sections of the model. In addition, we present a targeted mutant generation approach to reduce the number of redundant mutants. This directed approach focuses on mutating parameter values and gates that would have the greatest effect on the QML model based on the structure of the gates present in the quantum circuit.

Specifically, this paper makes the following contributions:
\begin{itemize}
    \item We define new mutation operations for feature mapping and the ansatz of QNNs.
    % , \textcolor{red}{related to the parameterized gates within}.
    \item We present a directed mutation generation approach for specifically crafting mutants to reduce the number of redundant mutant circuits.
    % \item \textcolor{red}{Our generated mutants are written to QASM files, for execution in any supported quantum simulator}.
    \item We demonstrate that our directed approach covers faulty scenarios in QML models that are not sufficiently addressed by previously defined mutation operations.
\end{itemize}

The rest of the paper is organized as follows. Section~\ref{sec:bg} provides a background on mutation testing and quantum machine learning. Section~\ref{sec:related} surveys related efforts in mutation testing of quantum computing as well as classical machine learning. Section~\ref{sec:method} describes our mutation testing framework for quantum machine learning models. Section~\ref{sec:result} presents the experimental results. Finally, Section~\ref{sec:conc} concludes the paper.

% \begin{itemize}
% \item Motivation - what is the objective (debugging, noise, fidelity, ...)
% \item State-of-the-art
% \item What is your hypothesis that we can do better
% \item List all possible mutations and see if there is a common denominator that covers all/most of them (e.g., bit flip in reliability or stuck-at-0 in manufacturing testing).
% \item Are there ny assumptions in the random sampling theory (e.g., in N-detect assuming is if each node is activated N times and N is sufficiently large)?
% \item Should we do random or directed generation of mutations (e.g., we start by flipping the output, and start figuring our layer/neuron-by layer/neuron, which changes will flip the output).
% \end{itemize}

\section{Background} \label{sec:bg}
In this section, we provide background on mutation testing (Section~\ref{sec:bg:mut}) and quantum machine learning (Section~\ref{sec:bg:ml}). 

\subsection{Mutation Testing} \label{sec:bg:mut}
In software domain, mutation testing is used to create faulty variants of the program under test, known as \textit{mutants}, ensuring that the test suite is capable of evaluating the program against the design specification~\cite{jia2011analysis}. These mutants are crafted using a \textit{mutation operation}, which introduces a fault into the program by altering the program in a specific manner. For quantum computing, mutation testing defines the quantum circuit as the program. Thus, mutation operations act on the components of the quantum circuit, such as gates. Prior work~\cite{mendiluze2021muskit, fortunato2022mutation} has defined five basic mutation operations for quantum circuits. These are defined as follows:
\begin{enumerate}
    \item \textit{\textit{Gate addition}} - adding a random gate into a location within the quantum circuit.
    \item \textit{\textit{Gate removal}} - removing a random gate from a location within the quantum circuit.
    \item \textit{\textit{Gate replacement}} - replacing a random gate from a quantum circuit with a semantically equivalent gate. This means that a two-qubit gate will be replaced by another two-qubit gate, and so on.
    \item \textit{\textit{Measurement addition}} - adding a measurement operator into the circuit.
    \item \textit{\textit{Measurement removal}} - removing a measurement operator from the circuit.
\end{enumerate}

\begin{figure}
    \centering
    \includegraphics[width=\linewidth]{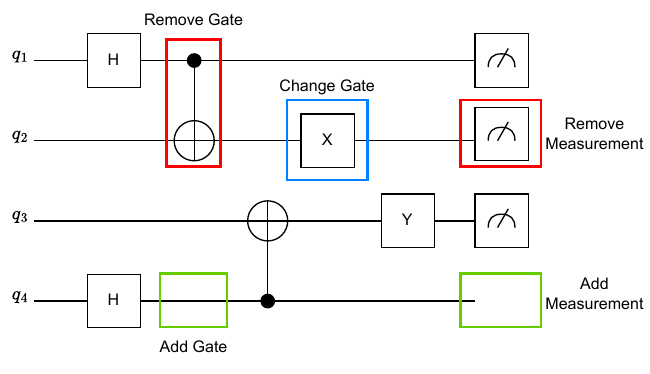}
    \caption{Example of mutation operations. The CNOT gate between qubits $q_1$ and $q_2$ is removed with a remove gate operation. A new one-qubit gate is added with the add gate operation after the first Hadamard gate on qubit $q_4$. The Pauli-X gate acting on qubit $q_2$ is changed to a semantically equivalent one-qubit gate on the same qubit and in the same location with the change gate operation. The measurement of $q_2$ is removed with the remove measurement operation. A new measurement is added at the end of the circuit on qubit $q_4$ with the add measurement operation. Note that typically one mutation is performed to generate the mutant.}
    \label{fig:mutex}
    \vspace{-0.2in}
\end{figure}

Figure~\ref{fig:mutex} showcases an example of these mutation operations in practice on a real quantum circuit. In this circuit, any gate can be changed to a semantically equivalent gate. This means that a one-qubit gate will be changed for a different one-qubit gate within the available gate set. Removing a gate will remove the specific gate from the circuit, while adding a gate will place a gate from the available gate set into any available position and qubit ordering possible. 

Mutation operations are also defined for classical machine learning models~\cite{ma2018deepmutation, wang2018detecting}. These mutation operations are used to insert faults into the model and to alter the data sample inputs. Example mutation operations include, but are not limited to:
\begin{enumerate}
    \item \textit{\textit{Layer addition}} - adding a layer into the model
    \item \textit{\textit{Layer deletion or deactivation}} - removing or deactivating a layer in the model
    \item \textit{\textit{Weight adjustment}} - adjusting the learned weight in a model, often by fuzzing against a known probability distribution, such as the Gaussian distribution.
\end{enumerate}
These mutation operations alter the model or the data to evaluate the quality of the test suite or to find potential areas of sensitivity in the model.

\subsection{Quantum Machine Learning} \label{sec:bg:ml}
\begin{figure}[htp]
    \centering
    \includegraphics[width=\linewidth]{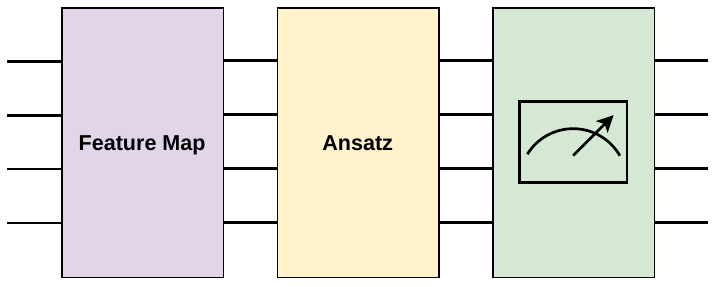}
    \caption{Components of a quantum neural network. The data is first embedded into the quantum circuit using a feature mapping. With the embedded data, the ansatz can process it, learning features and other relationships depending on the model objective. Finally, measurement of the quantum circuit produces the result of the model, such as a class label in classification tasks.}
    \label{fig:qnn}
\end{figure}

Classical machine learning trains a model to perform a task on a group of data. In the neural network family of machine learning models, the models contain layers performing computations. These layers contain weights, which are updated as part of the training process. QML also uses this same notion, with quantum circuits performing the computation. In QNNs, there are three components: the feature map, the ansatz, and measurement, as depicted in Figure~\ref{fig:qnn}. 
%We detail the operation of each below.

\subsubsection{Feature Mapping}
\begin{figure}[htp]
    \centering
    \includegraphics[width=\linewidth]{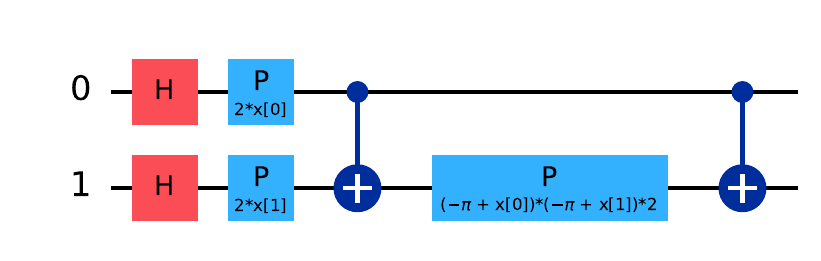}
    \caption{Example of two-qubit ZZ feature map. $x[0]$ and $x[1]$ are the values taken from the input data sample.}
    \label{fig:zzfm}
\end{figure}

The first component of QNNs is feature mapping, where input data samples are embedded into the quantum circuit. The embedding occurs by using values of the input data sample as parameters to parameterized gates in the chosen feature mapping. An example feature map is shown in Figure~\ref{fig:zzfm}. This is the ZZ feature map, which performs the second-order Pauli-Z evolution. Two features from the input data sample are encoded into the quantum circuit through the parameters of the P gates, taking the values of $x[0]$ and $x[1]$.

\subsubsection{Ansatz}
The second component of QNNs is the ansatz, which performs the actual computation of the model. This portion of the circuit consists of parameterized gates with entangling operations. The parameters of the gates act as the weights of the model, which are learned during training.

\subsubsection{Measurement}
Measurement operations are used at the end of the quantum circuit to obtain the resulting output. The specific output is the result of the task the model is performing.

\begin{figure}
    \centering
    \includegraphics[width=\linewidth]{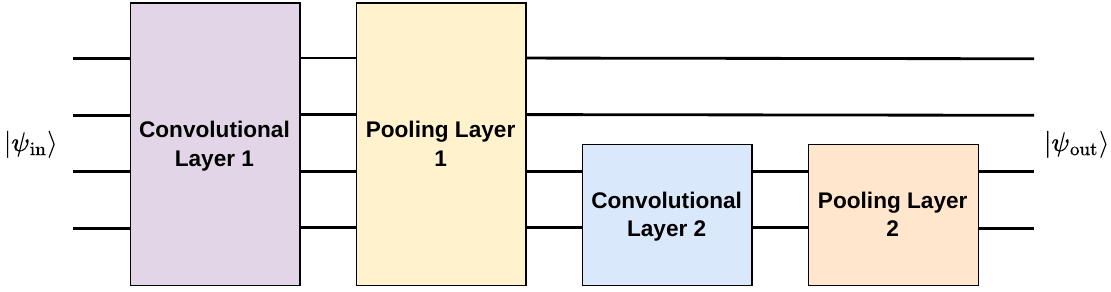}
    \caption{Typical structure of a quantum convolutional neural network (QCNN). It consists of a convolutional layer followed by a pooling layer. The next round of these pairs of layers reduces the number of qubits it operates over.}
    \label{fig:qcnn}
\end{figure}

The quantum circuit consisting of all three components produces output learned through the training process against a series of metrics, such as accuracy of predicted labels in classification tasks. This QNN structure has been adapted into variants of the QNN, such as the quantum convolutional neural network (QCNN), which classifies image-like data samples~\cite{cong2019quantum}. Figure~\ref{fig:qcnn} showcases the typical layer structure within the ansatz for QCNNs.

\section{Related Work} \label{sec:related}
In this section, we survey related efforts for mutation testing of quantum circuits (Section~\ref{sec:rel:qc}) as well as  classical machine learning models (Section~\ref{sec:rel:cml}).

\subsection{Mutation Testing of Quantum Circuits} \label{sec:rel:qc}
There are various prior effors establishing mutation testing and mutation operations for quantum circuits. Muskit is a mutation testing tool that defines the original mutation operations for adding, removing, and changing a gate in the quantum circuit~\cite{mendiluze2021muskit}, as described in Section~\ref{sec:bg:mut}. A given quantum circuit is mutated with the mutation operations in an exhaustive manner, meaning every possible mutation is considered with all gates in the given quantum circuit. Muskit uses two test oracles from other prior efforts~\cite{ali2021assessing, wang2021quito} to determine whether or not the mutant is killed. These two oracles are described below.
\begin{itemize}
    \item The \textit{Wrong Output Oracle (WOO)} examines all the output bitstrings produced by the executed mutants. If any bitstrings are incorrect given the output bitstrings of the original quantum circuit, then that mutant is killed.
    \item The \textit{Output Probability Oracle (OPO)} evaluates the output of the mutants if all output bitstrings are valid, testing if the probability distribution of the output bitstrings is the same as the original quantum circuit. Note that due to environmental and other factors, the exact counts of each output from multiple executions of the quantum circuit can vary slightly, even if the same circuit were executed. Thus, the output probability distributions are compared within a small tolerance, such as 5\%.
\end{itemize}

QMutPy is another mutation testing framework~\cite{fortunato2022mutation}, defining all five mutation operations detailed in Section~\ref{sec:bg:mut}. Diff patches are used to construct the mutation on the quantum circuit written in Qiskit. Other mutation testing approaches within quantum computing include search-based approaches, where the searches try to find equivalent mutants and remove them from testing to reduce redundant mutants~\cite{wang2022mutationbased}. Additional analysis of the mutation operations in the context of a variety of different quantum applications has also been carried out to potentially reduce redundant mutants occurring with the given mutation operations~\cite{mendiluzeusandizaga2025quantum}.

However, these prior efforts have practical limitations in extending to QML models. While these techniques focus on generalized mutation operations for a wide variety of quantum circuits, they are not specific enough for certain applications, allowing for different categories of mutations to potentially produce faulty models. Additionally, these frameworks produce many redundant mutants, increasing as the circuits grow in complexity. Therefore, for QML models, there needs to be additional mutation operations to cover additional mutation functionality present in these specific models and for efficient generation of the mutants.

\subsection{Mutation Testing of Classical Machine Learning Models} \label{sec:rel:cml}
Mutation testing frameworks in classical machine learning craft mutation operations that alter the datasets used, the training process, and the model itself to create a faulty program from all aspects. DeepMutation is a framework that focuses on mutating these three aspects in deep learning systems to evaluate the test data used~\cite{ma2018deepmutation}. To do so, the mutation operations are broken up into those affecting any aspect of the deep learning system during the actual training process and altering the model after training. Mutation operations for after training focus on altering properties of the model, such as the layers and weights. For example, the weights can be perturbed by adding noise sampled from the Gaussian distribution to them, or the layers can be removed or added.
% \begin{enumerate}
%     \item Gaussian fuzzing - use the Gaussian distribution to fuzz or perturb layer weights.
%     \item Weight shuffling - weights are shuffled.
%     \item Neuron effect block - block a neuron from affecting the following layer.
%     \item Neuron activation inverse - a neuron's activation status is inverted.
%     \item Neuron switch - within a layer, two neuron are switched.
%     \item Layer deactivation - a layer is deactivated.
%     \item Layer addition - a layer is added.
%     \item Activation function removal - an activation function is removed.
% \end{enumerate}
These mutations help to evaluate the test set and also to identify portions of the model that are sensitive to changes about the model.

Adversarial robustness is also a major concern in validation of machine learning models, and mutation testing is suited to be a useful tool in identifying these samples. Wang et al.~\cite{wang2018detecting} showcased how mutation testing can be utilized to detect potential adversarial samples that can cause a label change. Mutation operations are performed on the data itself, cataloging the resulting class label predicted by the trained model running on the mutated sample.  The classical mutation operations and mutation testing methods cannot be directly applied on QML models. To the best of our knowledge, there are no prior efforts in mutation testing of QML models.
\section{Mutation Testing of QML Models} \label{sec:method}
\begin{figure}[h]
    \centering
    \includegraphics[width=\linewidth]{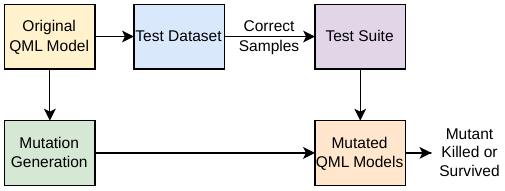}
    \caption{Overview of our mutation testing framework. The original trained model is tested on the test dataset, saving the correctly classified samples as the test suite. The original model is then mutated with the given mutation operations, producing mutants. These mutants are evaluated on the test suite, resulting in each mutant either killed or survived.}
    \label{fig:overview}
\end{figure}

Figure~\ref{fig:overview} presents an overview of this mutation testing workflow, taking the original QML model, mutating it with the mutation operations, then evaluating both the original model and the mutated models on the test oracle using metrics such as mutation score. We begin by establishing necessary preliminaries, including the available gate set and test oracles. Next, we present the new mutation operations for QML circuits, focusing on the feature map and the ansatz components. Finally, we discuss directed changes, influenced by the structure of gates in QNN circuits. 

\subsection{Available Gates and Simulators}
Part of the mutation operations previously defined and newly defined is changing a gate currently in the quantum circuit to a new gate. This new gate must come from an \textit{available gate set}, as it is possible that the device or simulator that executes the resulting quantum circuit may or may not have the capability to process all gates. 

When mutants are generated, we export mutants to QASM, a programming language designed to specify quantum circuits and their designs while being agnostic to a specific quantum software stack. This means that any quantum software stack can import a quantum circuit defined in QASM into their quantum circuit representation, if supported. 

\subsection{Test Suite and Evaluation Metrics} \label{sec:testsuite}
To examine the effectiveness of mutants produced by our framework, we must evaluate them against test oracles and score how well the test oracles performed at detecting mutated circuits through mutation score and survival rate.

\subsubsection{Test Suite}
To carry out the killing of mutants, we utilize testing data samples from the dataset to evaluate the mutated model. The original model first predicts the labels of the testing data samples. These predicted labels are compared against the true labels, splitting the testing samples into those the model predicted correctly and those predicted incorrectly. The correctly predicted samples are taken as the test suite. The mutated model is evaluated on this test suite of correctly predicted samples, killing any mutated model that no longer predicts the correct label. Predicting an incorrect label on a previously correct prediction indicates that the decision boundaries were altered in the model in the case of mutations to the ansatz components or that the decision boundaries are not robust enough to perturbations when under data sample mutations.

% The \textit{\textbf{Accuracy Oracle (AO)}} inspects the accuracy of a QML model on its produced output for a given task. For example, with classification tasks, this oracle examines if the accuracy of predicting the correct class labels for given data samples is equivalent to a tolerance between the original QML model and its weights versus the mutated QML model. The tolerance is needed to account for minor inconsistencies between multiple executions.

% \begin{figure*}[h]
%     \centering
%     \includegraphics[width=0.7\linewidth]{figs/framework.pdf}
%     \caption{Potential mutations on a QNN model. Mutation \textcircled{\scalebox{0.8}{1}} zeros a pair of feature samples $x_1$ and $x_2$. Mutation \textcircled{\scalebox{0.8}{2}} adds $\pi$ to parameters $t_1,t_2,t_3,$ and $t_4$ of a layer. Mutation \textcircled{\scalebox{0.8}{3}} changes a parameterized $R_Z$ gate to an $R_X$ gate, maintaining the parameter value $t_5$. {\color{red}TODO: UPDATE}}
%     \label{fig:mutations}
% \end{figure*}

\begin{figure*}
    \begin{subfigure}{0.245\linewidth}
        \centering
        \includegraphics[width=\linewidth]{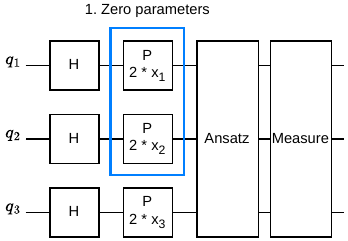}
        \caption{APC}
        \label{fig:apc}
    \end{subfigure}
    \begin{subfigure}{0.245\linewidth}
        \centering
        \includegraphics[width=\linewidth]{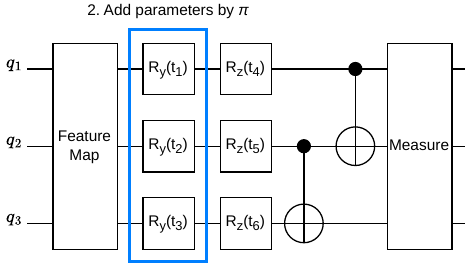}
        \caption{DFC}
        \label{fig:dfc}
    \end{subfigure}
    \begin{subfigure}{0.245\linewidth}
        \centering
        \includegraphics[width=\linewidth]{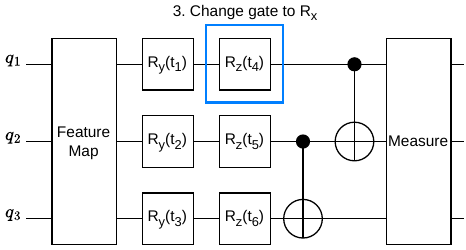}
        \caption{APGC}
        \label{fig:apgc}
    \end{subfigure}
    \begin{subfigure}{0.245\linewidth}
        \centering
        \includegraphics[width=\linewidth]{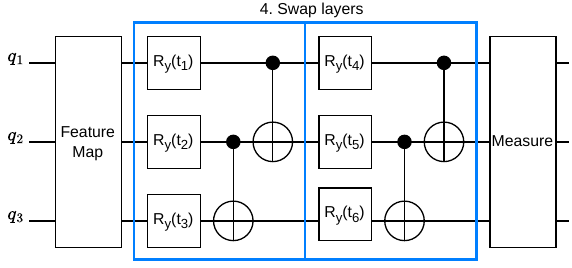}
        \caption{LS}
        \label{fig:ls}
    \end{subfigure}

    \begin{subfigure}{0.32\linewidth}
        \centering
        \includegraphics[width=\linewidth]{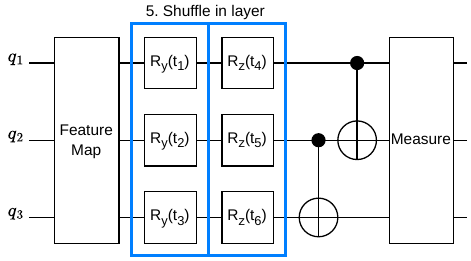}
        \caption{ILS}
        \label{fig:ils}
    \end{subfigure}
    \begin{subfigure}{0.32\linewidth}
        \centering
        \includegraphics[width=\linewidth]{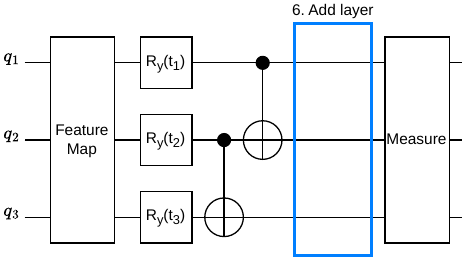}
        \caption{ALA}
        \label{fig:ala}
    \end{subfigure}
    \begin{subfigure}{0.32\linewidth}
        \centering
        \includegraphics[width=\linewidth]{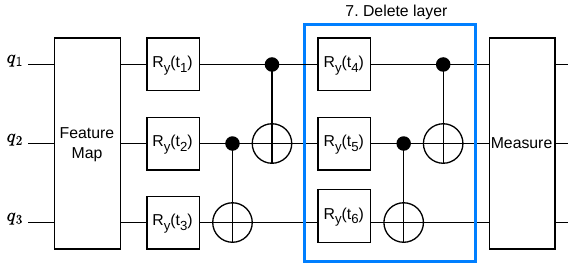}
        \caption{ALD}
        \label{fig:ald}
    \end{subfigure}
    \caption{Example operations for each of the seven defined mutation operations on QML models.}
    \label{fig:mutations}
    \vspace{-0.1in}
\end{figure*}

\subsubsection{Mutation Score}
Mutation score is the ratio of killed mutants over all total mutants generated. This can be formally defined as
\begin{equation} \label{eq:mutscore}
    MS=\frac{|K|}{|O|}\times 100\%,
\end{equation}
where $O$ is the set of all mutation operations generated and $K$ is the set of all killed mutants. 
% The survival rate is similar, measuring the ratio between the number of surviving mutants to the total number of mutants. It can be expressed as
% \begin{equation}
%     SR=\frac{|S|}{|O|}\times 100\%,
% \end{equation}
% where $S$ is the set of surviving mutation operations. 

\subsection{QML Mutation Operations}
We have devised seven new mutation operations to target the structure of QNN models more effectively compared to the mutation operations considered in prior work~\cite{mendiluze2021muskit, fortunato2022mutation} (see Section~\ref{sec:bg:mut}). 
% These new mutation operations are grouped into three categories: ansatz parameter changes, data sample feature changes, and ansatz parameterized gate changes. 
Within each category, often the mutation operation involves changing a value. A key challenge in generation of mutants is to ensure that redundant mutants are not generated. Thus, we propose directed efforts to both mutate reliably without using random values and to reduce the number of redundant mutants, often created from the commutativity of quantum circuits. An overview of each of the seven mutation operations is shown in Figure~\ref{fig:mutations}.

\subsubsection{Ansatz Parameter Change}
The ansatz consists of many $R_X$, $R_Y$, and $R_Z$ gates, parameterized by some $\theta$ value. The unitary matrix of these gates is a function valued by $\theta$. For example, $R_Y(\theta)$ has the unitary matrix of
$$\begin{bmatrix}
    \cos(\frac{\theta}{2})&-\sin(\frac{\theta}{2})\\
    \sin(\frac{\theta}{2}) & \cos(\frac{\theta}{2})
\end{bmatrix},$$
meaning that the learned parameter for this gate in the ansatz will be related to trigonometric functions and define the exact gate operations differently. We use the notion from classical machine learning of mutating the weights of a layer~\cite{ma2018deepmutation}, mutating a collection of gates' parameters acting as a layer at once, resulting in the \textbf{\textit{ansatz parameter change (APC)}} mutation operation.

The definition of a layer is dependent upon the design of the ansatz. For example, a layer can consist of all parameterized gates inside of a repetitive block. This can be further refined by analyzing a series of one-qubit gates performing on multiple qubits at the same time evolution within a repetitive block.

The changes to the layer parameters are performed with four different categories of operations. These operations are:
\begin{itemize}
    \item \textit{Zeroing} of the parameter values;
    \item \textit{Sign flipping} the parameter values;
    \item \textit{Addition of a perturbation} to the parameter values; and
    \item \textit{Scaling of a perturbation} to the parameter values.
\end{itemize}
Both zeroing and sign flipping are fixed operations, meaning that there is strictly one operation associated with the category. On the other hand, both addition and scaling require an additional value. 

A straightforward way to get this additional value is taking a random sample from a known probability distribution, such as the Gaussian distribution. While this does change the parameter value, often times it is not sufficient enough of a change based on the manner in which the parameter value is evaluated in the actual unitary matrix of the gate. Thus, we use known radians such as $\pi/2$ and $\pi$ to ensure that the phase is shifted. Additionally, this restricts the number of potential radians to check between $[-\pi, \pi]$. $\theta=2\pi$ is not necessary to check as it is covered by $\theta=0$, as evidenced by evaluating $\cos(2\pi)=\cos(0)$.

\subsubsection{Data Sample Feature Changes}
The features of the input data sample are used as parameters in the feature mapping section of the QNN. Therefore, we can directly alter features of the input data sample to cause a mutation in the \textbf{\textit{data sample feature change (DFC)}} mutation operation. Effective mutations in feature map circuits will involve breaking entanglement and symmetry between parameterized gates loading the data sample features into the circuit. Thus, this can be done by mutating several different features at once. For a pair of features, we define several different operations to mutate the feature value serving as the parameter to a feature map:
\begin{itemize}
    \item \textit{Addition} of a fixed value;
    \item \textit{Multiplication} of a fixed value;
    \item \textit{Sign flipping}; and
    \item \textit{One minus} of parameter value.
\end{itemize}

These can be combined such that one parameter value in the pair is mutated through one operation and the other parameter value either the same or different operation. Like with the ansatz parameter change, the value to scale in the addition and multiplication operations can be sampled from a probability distribution, or targeted based on the gates. In the targeting scheme, radians between $[-\pi,\pi]$ are chosen to potentially break entanglement and symmetry within the feature map that loads the data for the QNN ansatz to predict a label from.

For images, each pixel is a feature that is encoded with the feature map. Therefore, the classical data mutations on images are beneficial to potentially break entanglement and symmetry within the feature map from the original data sample as they affect many pixels in the image at once. These operations include cropping the image, rotating the image, horizontally flipping the image, and vertically flipping the image. 

\subsubsection{Ansatz Parameterized Gate Change}
In the \textbf{\textit{ansatz parameterized gate change (APGC)}} mutation operation, we change a parameterized gate to another parameterized gate, maintaining the original parameter value. To refine and direct this change, we analyzed the potential set of parameterized gates that are often used in QNNs. Popular one-qubit parameterized gates include $R_X(\theta),R_Y(\theta),$ and $R_Z(\theta)$, each of which has a different impact on the qubit value the gate is operating on. $R_X(\theta)$ and $R_Y(\theta)$ causes the amplitude to change, whereas $R_Z(\theta)$ adjusts the phase shift. We target parameterized gate changes so that the original parameterized gate in the original quantum circuit is changed to a parameterized gate with a different impact on the qubit value.

\subsubsection{Layer Shuffling}

The ansatz features layers, which are groupings of parameterized rotation gates followed by entangling gates. We mutate the structure of the layers in the \textbf{\textit{layer shuffling (LS)}} mutation operation by shuffling the layers in the model. Depending on the model, it is possible to produce incompetent mutants, or mutants that have syntax or similar unrecoverable errors. These mutants are noted separately from the killed and surviving mutants and are not counted in the mutation score.

Each layer consists of a series of parameterized gates followed by entangling gates. We switch two layers around in the model for this mutation operation. Often, switching entire layers causes all test samples to fail, and is thus not a nuanced enough fault in the model. To find the smaller changes that cause minor faults that could potentially go undetected otherwise, we also implement shuffling between the gates within a layer, in several different configurations. One is switching the ordering of the rotation gates and the entangling gates, such that the entangling gates are placed first in the layer followed by the rotation gates. If there are more than one rotation gate set per layer, then this shuffling phenomenon can stretch to all possible swaps between the gate sets in the layer itself. Note that these swaps are done one at a time, meaning that the gates inside of only one layer are swapped per mutant.

\subsubsection{In-layer Shuffling}

The LS mutation operation can be extended to shuffling components inside of a layer, resulting in the \textbf{\textit{in-layer shuffling (ILS)}} mutation operation. The layers often contain several components within them, such as parameterized gates followed by entangling gates. In this mutation operation, the parameterized gates are swapped with the entangling gates, meaning that the mutated layer in this example is the entangling gates followed by the parameterized gates. This notion can be extended to any blocks that make up a single layer in the ansatz, swapping their orientations in the mutated model.

% \subsubsection{Entangler Shuffling}
% {\color{red}
% The entangling gates in each layer are an instrumental part of the structure of the QML model to learn features effectively. These gates create or enhances entanglement between two qubits. In the context of the QML models, this entanglement allows for advanced learning of features that is otherwise not possible in classical computing. Therefore, the \textbf{\textit{entangler shuffling (ES)}} mutation operation aims to break the entanglement patterns used to learn the features by swapping the gates around.

% To reduce on duplicates due to commutativity between CNOTs, we utilize patterns. For smaller qubit models, we can exhaustively do all positions of the CNOT gates effectively per layer. However, for as the qubits grows, this becomes challenging, as the total number of permutations grows, with the number resulting in duplicates due to commutativity also increasing. For larger qubits, we use a shuffling pattern instead, where the linear chain is reversed in order. Any single gate changes that is done on the way to achieve the reverse chain is used as a pattern. This means that if there is a 5 qubit model, there are gates (0, 1, 2, 3). We shuffle in the pattern of (1, 0, 2, 3), (2, 1, 0, 3) and (3, 2, 1, 0).
% }

\subsubsection{Ansatz Layer Addition}

We expand on the gate change mutation operation previously defined to add additional layers into the models. We add an additional layer to the model with the \textbf{\textit{ansatz layer addition (ALA)}} mutation operation. This layer is added to any possible spot after existing layers in the models.

As this is adding a new layer into the model, the parameterized rotation gates will have new parameters contributing to the model. These parameters must be set prior to the resulting mutated model being used for predictions, otherwise this mutation operation would only result in incompetent mutants. We utilize both random initialization of the weights and copying of another layer's weights to populate the parameters to create competent mutants.

\subsubsection{Ansatz Layer Deletion}

Similar to the layer addition, we expand gate deletion mutation operations to a layer, resulting in the \textbf{\textit{ansatz layer deletion (ALD)}} mutation operation. Specifically, we target the entire layer, the rotation gates only, and the entangling gates only. We delete all gates consisting of a specific layer as the mutation operation. This results in a portion of the learned weights also being deleted that were associated with the deleted layer. 

These seven new mutation operations result in mutants that cover structural changes of the QML models.

\section{Experimental Results} \label{sec:result}
In this section, we evaluate the effectiveness of our mutation testing framework. We first describe the experimental setup. Next, we present results of mutating several different models and datasets, analyzing their mutation scores, generation times, evaluation times, and the reduction in redundant mutants. 

\subsection{Experimental Setup}
We use Python v3.13.5, Qiskit v2.2.3~\cite{javadi-abhari2024quantum}, and Qiskit Machine Learning v0.9.0~\cite{sahin2025qiskit}. Mutants are exported to QASM v3~\cite{cross2022openqasm} for best compatibility with parameter definitions. For execution, we test on the classical simulators provided by Qiskit-Aer v0.17.2. All experiments are run on an Intel i9 13900K CPU. In addition to the newly defined mutation operations, we also perform experiments with the existing mutation operators to evaluate the different facets of the QML models. Note that while we are using the mutation operations as defined in previous work~\cite{mendiluze2021muskit, fortunato2022mutation}, we implement the mutation operations in our framework to enable a fair comparison. 

\subsubsection{Datasets}
To evaluate our mutation testing framework, we utilize different feature maps, ansatzes, and datasets. This ensures that a variety of different circuit architectures and trained performances are examined. The datasets we utilize are tabular and image classification datasets:
\begin{itemize}
    \item \textbf{\textit{Iris}}~\cite{iris}, a tabular classification dataset with 4 features and 3 class labels;
    \item \textbf{\textit{Wine Quality}}~\cite{wine_quality}, a tabular classification dataset with 11 features and 11 class labels;
    \item \textbf{\textit{Wisconsin Breast Cancer (BC)}}~\cite{breast_cancer}, a tabular classification dataset with 30 features and 2 class labels;
    \item \textbf{\textit{MNIST}}~\cite{lecun1998mnist}, an image classification dataset of handwritten digits with 10 class labels;
    \item \textbf{\textit{FashionMNIST (FMNIST)}}~\cite{xiao2017fashionmnist}, an image classification dataset of fashion items with 10 class labels; and
    \item \textbf{\textit{Kuzushiji-MNIST (KMNIST)}}~\cite{clanuwat2018deep}, an image classification dataset of Hiragana characters with 10 class labels.
\end{itemize}

\subsubsection{QML Models}
Data samples from these datasets can be feature mapped into the quantum circuit of the QNN in three different ways. 
% \textcolor{red}{Each mapping has its own qubits to features ratio, and for some data types, this can restrict the mappings available.} 

\begin{itemize}
    \item \textbf{\textit{ZFeatureMap (ZFM)}} is the first order Pauli-Z evolution where $n$ features are mapped to $n$ qubits;
    \item \textbf{\textit{ZZFeatureMap (ZZFM)}} is the second order Pauli-Z evolution where $n$ features are mapped to $n$ qubits; and
    \item \textbf{\textit{Amplitude Embedding (AE)}} maps $2^n$ features into $n$ qubits through state preparation of the statevectors.
\end{itemize}
Tabular classification data typically features low numbers of features or features that can be easily reduced through a dimensionality reduction technique such as principal component analysis (PCA), making ZFM and ZZFM ideal choices. With images, each pixel is a feature, making these two feature maps unfeasible for use. Therefore, amplitude embedding is a better fit to allow for the image to be of sufficient size while reducing the number of qubits. We resize the images from the image datasets from the original size of $28\times28$ to $16\times16$ to map into 8 qubits.

Like with feature maps, there are several different ansatz that can be used. Each ansatz is characteristic with the notion of layers in classical machine learning, as well as consisting of blocks that can be repeated multiple times for longer range entanglements. The three ansatzes we focus are:
\begin{itemize}
    \item \textbf{\textit{RealAmplitudes (RA)}}, a strongly entangling ansatz featuring parameterized $R_Y$ gates;
    \item \textbf{\textit{EfficientSU2 (SU2)}}, a two local ansatz featuring parameterized $R_Y$ and $R_Z$ gates; and
    \item \textbf{\textit{QCNN}}~\cite{cong2019quantum},  quantum convolutional neural network with convolutional and pooling layers for image classification.
\end{itemize}
Note that as QCNN is primarily focused on image classification tasks, we will not use this ansatz for the tabular classification datasets. Additionally, the QCNN performs binary classification, so each image dataset is split into classifying class label 0 versus the other class labels in a one versus rest classification objective. 
% An example of the ZFM and SU2 QNN for the Iris dataset is shown in Figure~\ref{fig:qnncirc}.

Each dataset model is run on a 20 samples from their respective test sets. These 20 samples are run through the original model first, gathering which samples are predicted correctly according to the true class labels. The samples that the original model predicted correctly are gathered as the test suite for any mutated model from the corresponding original model. For consistency, the same test suite is used by both our mutation operations and prior mutation operations.

% \begin{figure*}[h]
%     \centering
%     \includegraphics[width=\linewidth]{figs/iris_z_su_circuit.pdf}
%     \caption{A four-qubit QNN model with the ZFM for embedding data and SU2 for the ansatz. This QNN model is used for the Iris dataset with an additional layer, and is extended to 8 qubits and 3 repeating layers for Wine and BC. The barriers represent the designations between the feature map and the individual layers within the ansatz.}
%     \label{fig:qnncirc}
% \end{figure*}

\begin{table*}[h]
    \centering
    \caption{Our mutation operations. Entries for the mutation operations are killed/total for the respective mutation operation.}
    \label{tab:ours}
    \begin{tabular}{ccccccccccccc}
        \toprule
        \textbf{Dataset} & \textbf{FM} & \textbf{Ansatz} & \textbf{MS (\%)} & \textbf{APC} & \textbf{DFC} & \textbf{APGC} & \textbf{LS} & \textbf{ILS} & \textbf{ALA} & \textbf{ALD} & \textbf{Gen. (s)} & \textbf{Eval. (s)}\\
        \midrule
        \multirow{4}*{Iris} & \multirow{2}*{ZFM} & RA & 71.59 & 139/240 & 162/240 & 729/960 & 24/36 & 31/36 & 29/36 & 20/36 & 0.0258 & 11.8541\\
        &  & SU2 & 56.38 & 360/380 & 247/380 & 1769/3040 & 29/57 & 131/285 & 31/57 & 47/57 & 0.0733 & 30.9375\\
        \cmidrule{2-13}
        & \multirow{2}*{ZZFM} & RA & 65.06 & 214/340 & 209/340 & 910/1360 & 35/51 & 31/51 & 29/51 & 32/51 & 0.0282 & 22.7359\\
        &  & SU2 & 80.92 & 312/320 & 218/320 & 2420/2560 & 29/48 & 124/240 & 32/48 & 24/48 & 0.0672 & 29.1214\\
        \midrule
        \multirow{4}*{Wine} & \multirow{2}*{ZFM} & RA & 83.09 & 99/360 & 231/360 & 1351/1440 & 9/27 & 15/27 & 14/27 & 16/27 & 0.0788 & 28.4071\\
        &  & SU2 & 90.19 & 334/600 & 371/600 & 4800/4800 & 31/45 & 133/225 & 32/45 & 35/45 & 0.2313 & 78.7628\\
        \cmidrule{2-13}
         & \multirow{2}*{ZZFM} & RA & 88.20 & 105/320 & 209/320 & 1264/1280 & 17/24 & 13/24 & 16/24 & 13/24 & 0.0843 & 28.8146\\
         &  & SU2 & 91.86 & 53/80 & 54/80 & 640/640 & 4/6 & 17/30 & 5/6 & 6/6 & 0.2107 & 13.9383\\
        \midrule
        \multirow{4}*{BC} & \multirow{2}*{ZFM} & RA & 75.08 & 112/640 & 106/640 & 2560/2560 & 5/48 & 1/48 & 0/48 & 3/48 & 0.1638 & 50.5736\\
         &  & SU2 & 80.20 & 178/640 & 117/640 & 5118/5120 & 2/48 & 16/240 & 9/48 & 1/48 & 0.2370 & 85.7884\\
         \cmidrule{2-13}
         & \multirow{2}*{ZZFM} & RA & 88.61 & 112/560 & 491/560 & 2240/2240 & 10/42 & 4/42 & 18/42 & 3/42 & 0.0756 & 49.2563\\
         &  & SU2 & 86.79 & 208/440 & 214/440 & 3520/3520 & 13/33 & 68/165 & 17/33 & 8/33 & 0.2165 & 71.5143\\
         \midrule
         MNIST & AE & QCNN & 93.13 & 139/1260 & 20/48 & 5880/5880 & 8628/8628 & 324/324 & 504/504 & 72/72 & 0.9267 & 306.6400\\
         \midrule
         FMNIST & AE & QCNN & 94.06 & 343/1470 & 24/56 & 6860/6860 & 10066/10066 & 378/378 & 588/588 & 84/84 & 0.6054 & 349.9791\\
         \midrule
         KMNIST & AE & QCNN & 92.84 & 80/1050 & 13/40 & 4900/4900 & 7190/7190 & 270/270 & 420/420 & 60/60 & 0.7633 & 285.8274\\
        \bottomrule
    \end{tabular}
\end{table*}

\begin{table*}[h]
    \centering
    \caption{Existing mutation operations~\cite{mendiluze2021muskit, fortunato2022mutation} on each dataset and model configuration. Entries for the mutation operations are killed/total for the respective mutation operation. These mutation operations were implemented in our framework for best comparison and to have the same test suite for each respective dataset and model configuration as Table~\ref{tab:ours}.}
    \label{tab:sota}
    \begin{tabular}{ccccccccc}
        \toprule
        \textbf{Dataset} & \textbf{FM} & \textbf{Ansatz} & \textbf{MS (\%)} & \textbf{Add} & \textbf{Delete} & \textbf{Change} & \textbf{Gen. (s)} & \textbf{Eval. (s)} \\
        \midrule
        \multirow{4}*{Iris} & \multirow{2}*{ZFM} & RA & 35.92 & 4808/13500 & 98/300 & 283/648 & 0.2826 & 96.1708\\
        &  & SU2 & 33.00 & 11561/35055 & 159/779 & 443/1026 & 0.8251 & 267.7039\\
        \cmidrule{2-9}
         & \multirow{2}*{ZZFM} & RA & 34.56 & 6581/19125 & 100/425 & 393/918 & 0.2880 & 167.8519\\
         & & SU2 & 29.73 & 8826/29520 & 100/656 & 301/864 & 0.7684 & 240.1785\\ 
        \midrule
        \multirow{4}*{Wine} & \multirow{2}*{ZFM} & RA & 38.04 & 8186/21465 & 128/477 & 465/1134 & 1.1302 & 307.0147\\
        &  & SU2 & 40.86 & 23468/57375 & 379/1275 & 888/1890 & 2.9643 & 882.5387\\
         \cmidrule{2-9}
         & \multirow{2}*{ZZFM} & RA & 57.03 & 10863/19080 & 239/424 & 597/1008 & 1.1092 & 307.2850\\
         &  & SU2 & 44.96 & 3439/7650 & 60/170 & 130/252 & 2.9260 & 142.8110\\
        \midrule
        \multirow{4}*{BC} & \multirow{2}*{ZFM} & RA & 4.31 & 1672/38160 & 8/848 & 88/2016 & 1.1907 & 560.3533\\
        &  & SU2 & 4.58 & 2853/61200 & 8/1360 & 97/2016 & 2.9647 & 933.7876\\
        \cmidrule{2-9}
        & \multirow{2}*{ZZFM} & RA & 10.81 & 3582/33390 & 49/742 & 248/1764 & 1.0898 & 542.4357\\
         &  & SU2 & 26.45 & 11050/42075 & 169/935 & 525/1386 & 3.0816 & 723.8027\\
         \midrule
         MNIST & AE & QCNN & 100.0 & 83160/83160 & 1848/1848 & 4032/4032 & 7.8500 & 1744.4850\\
         \midrule
         FMNIST & AE & QCNN & 100.0 & 97020/97020 & 2156/2156 & 4704/4704 & 7.8048 & 2249.7314\\
         \midrule
         KMNIST & AE & QCNN & 100.0 & 69300/69300 & 1540/1540 & 3360/3360 & 8.4427 & 1495.9733\\
        \bottomrule
    \end{tabular}
\end{table*}

\subsection{Comparison of Mutant Generation Time} \label{sec:time}

The time to generate a mutant depends on the mutation operation and the number of gates it must change. However, this time is still under one second for our mutation operations with each dataset and model configuration, as shown in Table~\ref{tab:ours}. For the Iris dataset, the generation time averages to 0.00025 seconds per mutant generated, with 132 mutants created for the RA ansatz models and 244 created for the SU2 models. In comparison, the prior mutation operations displayed in Table~\ref{tab:sota} average to 0.00032 seconds per mutant. The Wine and BC datasets utilize 8 qubits compared to Iris' 4 qubits, resulting in more gates and more mutants. For example, with Wine, the generation time averages to 0.00044 seconds per mutant and the prior mutation operation averages 0.00058 seconds per mutant.

QCNN has a similar trend as the structural mutants reduce the space that must be parsed through to find the space to carry out a gate level mutation. This results in an average generation time per mutant across all three datasets of 0.00055 seconds per mutant, with the prior mutation operations averaging 0.00108 seconds per mutant.

\subsection{Comparison of Evaluation Time}

By using the Qiskit Machine Learning framework to run the mutated circuits, we are able to use efficient simulation to evaluate  the mutants against the test suite. If a specific model runs more efficiently under a different simulation backend, the QASM exports enable the mutated circuits to be reconstructed in the platform of choice. Simulating the circuits under evaluation is the longest portion of the mutation testing framework, as each mutated circuit generated must be simulated on the test suite samples to acquire a predicted class. Unlike classical machine learning where a model only needs to be run once, the result from the quantum model can vary due to noise and other quantum phenomenon. Therefore, these models are run with multiple different shots. For consistency, each model, including the original, is run under 1,000 shots, with the resulting label the most occurring result.

For the QNN models, evaluation time varied by model configuration and mutation operation. The SU2 models took longer than the RA models to run, displayed in Table~\ref{tab:ours}, as they have more gates in a layer compared to RA. The 8 qubit QNN models for Wine and BC also took longer to run as there are more qubits and gates to simulate compared to 4 qubits with Iris. Additionally, some mutation operations insert more gates than others, which can affect overall evaluation time, such as ALA inserting many gates consisting of one layer for a single mutant requiring more time to simulate. For example, Wine averaged 0.1551 seconds per mutant on the new mutation operations compared to 0.1234 seconds per mutant with the prior mutation operations. The QCNN models all took around the same time to execute, averaging at 0.2255 seconds per mutant, as each QCNN model used the same structure with different datasets. The QCNN models also had more gates to evaluate compared to the QNN models, resulting in an increase in the time required to simulate. The prior mutation operations in Table~\ref{tab:sota} averaged an evaluation time of 0.2466 seconds per mutant generated. Like with the QNNs, the evaluation time is dependent on how many and which gates are mutated with respect to the original circuit.

\subsection{Comparison of Mutation Scores}

\begin{figure}[htp]
\vspace{-0.1in}
    \begin{subfigure}{\linewidth}
        \centering
        \includegraphics[width=\linewidth]{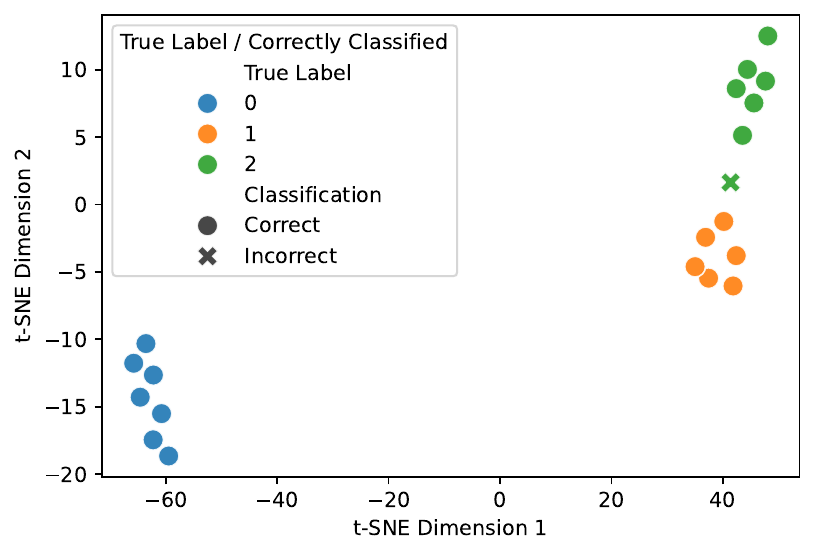}
        \caption{Original model and predictions. Correct samples are the test suite.}
        \label{fig:tsne-orig}    
    \end{subfigure}
    
    \begin{subfigure}{\linewidth}
        \centering
        \includegraphics[width=\linewidth]{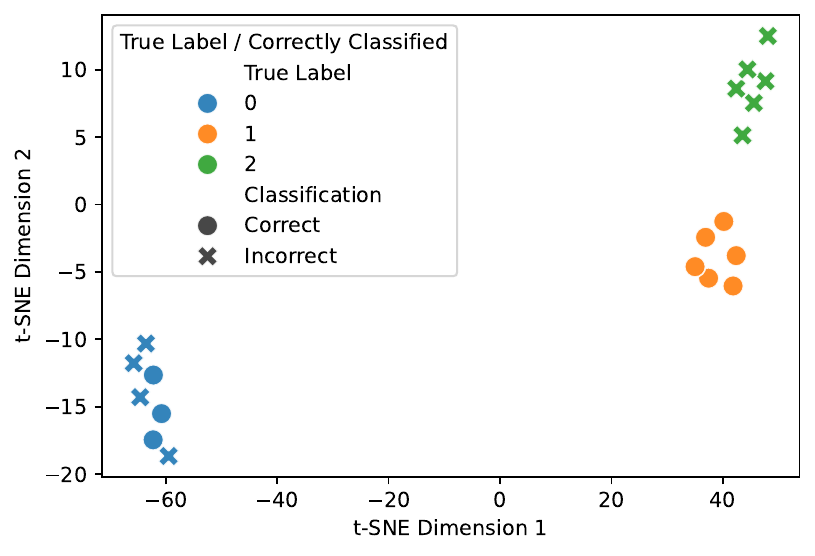}
        \caption{APGC mutation in the first ansatz layer.}
        \label{fig:tsne-param}       
    \end{subfigure}
    \caption{t-SNE of (a) original model and (b) APGC mutated model for the Iris dataset with ZFM and SU2.}
    \label{fig:tsne}
\end{figure}

The mutation score (Equation~\ref{eq:mutscore}) is the ratio of killed mutants to total mutants. For each model configuration and dataset, our mutation operations are able to be detected as mutants and killed by the test suite, as shown in Table~\ref{tab:ours}. In general, the greater the resulting change on the model  causing large changes in the decision boundaries, the more mutants are killed due to the sample being predicted incorrectly. An example of how a mutation operation affects the original model's decision boundaries is shown in Figure~\ref{fig:tsne}, where t-distributed stochastic neighborhood embedding (t-SNE)~\cite{maaten2008visualizing} is used to visualize the samples and their true label versus the model's prediction. In the case of the APGC mutated model, the effect of the mutation operation on the decision bounds in causing misclassifications is visible. We evaluate the mutation score for each mutation operation below.

\subsubsection{APC}
The APC mutation operation has varying mutation scores across different model configurations and datasets. The Iris QNNs had the best mutation score average of 78.27\%, which indicates that the weight changes in a layer were successful in having the model mispredict the data sample. The Wine and BC QNNs were more robust to the weight changes, with average mutation scores for APC of 45.56\% and 28.15\%, respectively. These models were more robust to the mutation operation, meaning that the specific layer weights only minimally affect the resulting decision boundaries. Additionally, depending on what the weight was changed to, it can potentially produce an equivalent effect to the original weight values, only modifying the decision boundaries slightly which may not be enough to have a sample be classified incorrectly. The QCNN models also show a similar trend, with an average mutation score of 13.99\% across all three datasets for the APC mutation operation. 

\subsubsection{DFC}
DFC also has varying mutation scores across different model types. The QNN models averaged 57.67\% mutation score, while QCNN averaged 39.01\%. This showcases that some mutations of the data samples are able to be predicted as the correct class, however, others cause the model in its trained ansatz to classify the mutated sample incorrectly. As these mutations affect the data samples directly and not the ansatz which processes the data, these mutants are useful to determine how robust the model's decision boundaries are when the data samples are noisy or adversarially crafted to get the model to predict incorrectly. The QCNN is more robust against these data sample mutations, able to still predict the correct class at a higher rate despite the original image being mutated, such as rotated or flipped. The QNNs, on the other hand, are more sensitive to changes to the data samples, resulting in more samples being predicted incorrectly and consequently being killed.

\subsubsection{APGC}
APGC was the most effective operation in killing data samples for all datasets and model configurations, with an average mutation score of 90.68\% for the QNN models and 100.00\% for the QCNN models. This is due to the parameterized rotation gates in a single layer changing to a different, semantically equivalent gate. The resulting unitary matrices tied to each gate are thus different, resulting in at times drastically different results. Depending on the position in the layer, this can have cascading effects, where the resulting error from the mutation operation grows, culminating in highly inaccurate decision boundaries to the original model. 

\subsubsection{LS and ILS}
For some models, LS and ILS were highly effective at killing mutants. QCNN had 100.00\% mutation score for both LS and ILS, while QNN averaged 47.01\% and 44.13\% for LS and ILS, respectively. The QCNN models successfully killed all mutants for LS as the layer shuffling breaks the structure of the QCNN that is reliant on, with the convolutional layers processing the data on a specific number of qubits and the pool layers reducing the number of qubits to operate on. When a pool layer is moved to the beginning of the model ansatz, for example, this immediately reduces the data to 2 qubits. Any convolutional layers after this reduction that attempt to learn features over a greater number of qubits will have broken learning, resulting in mutated decision boundaries. ILS causes the blocks inside of the layer to be swapped, also breaking learned features and reflecting in the changed decision boundaries. 

For the QNN models, as the layers are repeated in terms of the gates only differing by weight values, it is possible that layers can have similar weights, meaning that LS does not affect the decision boundaries and results in the mutant not being killed. ILS can break up some learned features and entanglement patterns, however, the minor architecture change may not affect the final processing of the sample to change the decision boundaries enough to cause the model to make an incorrect prediction.

\subsubsection{ALA and ALD}
Other highly effective mutation operation at killing mutants were ALA and ALD. In these mutation operations, the full architecture of the model was changed in the layers present to process the data through the ansatz. Across the QCNNs, each had a mutation score of 100.00\% for both ALA and ALD. This is due to the layers being highly structured and reducing the information to smaller qubits. If a pool layer is added or removed in the QCNN at any position, the data being reduced into smaller qubits is altered depending on the mutation operation specifically. In the case of layer addition, the data can be further reduced from normal or can attempt to reduce data that has already been reduced and is no longer affecting the model. Similarly, with deletion, the removal of pool layers can cause the data to not be reduced properly, breaking the required structure. A similar effect is produced with the addition and removal of convolutional layers, where the processing on the layer's respective number of qubits is either removed entirely or additionally added. This can have drastic effects on the processing as these gates can operate on qubits that have already been reduced via pool layers. 

The QNN models had reduced mutation scores compared to QCNN, with an average mutation score of 53.71\% for ALA and 48.47\% for ALD. Because each layer in the QNN model is the same unlike in QCNN, the addition or deletion of a layer may not cause the model to produce a different, incorrect prediction. This indicates that the layer that was either added or deleted has a minor impact on the resulting decision boundary. In some cases, this is present for many of the mutants created using ALA or ALD, such as with the BC QNNs. In each of the four corresponding model configurations, ALA and ALD did not successfully kill a majority of the mutants, and with ZFM and RA, failed to kill any with ALA. This is indicative of the repeated layer or the deleted layer not being able to change the decision boundaries to cause the model to make differing predictions from the original model.

\subsubsection{Prior Mutation Operations}
As shown in Table~\ref{tab:sota}, the prior mutation operations had varying levels of successfully being killed. The tabular dataset models were more robust to the mutations, with an average mutation score of 30.02\%. On the other hand, the QCNN models successfully killed all mutations. This is likely due to the sensitivity of the models. In QCNN models, the architecture is set to perform a convolutional layer then a pooling layer to condense the learned features into a smaller number of qubits. Because these qubits cannot be dynamically reassigned, gates can still be added to the circuit on qubits that are not part of the reduced range, resulting in a major architecture shift. Additionally, within the layers, there are blocks of parameterized and entangling gates that operate on specific qubits, where any deviation from these qubits, such as addition of a gate on a qubit not involved in those computations, affects the resulting entanglement and overall processing of the model.

\subsection{Reduction of Redundant Mutants}
%large after refering to lower
The prior mutation operations as shown in Table~\ref{tab:sota} produce varying levels of mutation scores, varying for killing all mutants in QCNN models to smaller percentages with the QNN models. For both model types, there are many mutants being generated, particularly with the add gate operation. For QNNs, this can generate as many as 61,200 mutants, whereas for QCNN it can generate up to 97,020. With this many mutants, more generation and evaluation time is required, as discussed  in Section~\ref{sec:time}. Another key aspect is that many of these mutants produce equivalent faulty behavior, thus causing redundant mutants that do not need to be executed. 

Our mutation operations, showcased in Table~\ref{tab:ours}, provide a directed approach to mutating the structure of the QML models to reduce the redundant mutants being generated and explore additional fault patterns not observed with prior mutation operations. With the directed approach targeting specific structural changes, these mutation operations can generate up to 6,784 mutants (BC with ZFM and SU2), a reduction of 9.5 times, down from 64,576 under the same model configuration with prior mutation operations. Additionally, these mutation operations are designed to limit the number of redundant mutants being generated due to commutativity within the circuit. For example, if a selection of gates on a group of qubits were added before and after an existing gate operating on a different qubit, that two mutants are not generated as the inserted gates would occur at the same time evolution as the original gate.

% With the new mutation operations, the fault patterns are easier to observe with less mutants under high-level architectural changes compared to low-level circuit changes. The low-level changes often require many redundant mutants to be generated to observe specific fault patterns, if possible. We can easily see the decision boundary changes as a result of the mutation operations by plotting the features and the resulting classification using t-distributed stochastic neighborhood embedding (t-SNE)~\cite{maaten2008visualizing}. 
% add ref to table/works
In prior mutation operations, the add gate operation produces the majority of the mutants, as every possible gate is added into places within the quantum circuit. This can produce equivalent mutants with commutativity. Additionally, small gate additions may not cause the model to deviate enough to flip decision boundaries from correct prediction to incorrect prediction. This indicates that the decision boundaries originally trained by the model maintain some robustness to them to account for small noise or perturbations within the model.
% \textcolor{red}{which the addition of a single gate in the entire ansatz can result in}.

\section{Conclusion} \label{sec:conc}
This paper proposed a new mutation testing framework, specifically targeting QNNs to ensure that changes to the model's output predictions are detected. We present seven new mutation operations which are able to find faults that would not be exposed with prior mutation operations. Specifically, these mutation operations target mutations to the data samples and the ansatz, resulting in mutants able to mutate the decision boundaries of the model effectively. Our directed approach also reduces the number of overall and redundant mutants required to observe faulty behavior, enabling efficient testing of quantum neural networks.
 
 % We can successfully detect mutants affecting the ansatz parameter weights, mutations to the features of the input data samples, and changes to the parameterized gates of the ansatz. In addition, with the mutations to the input data samples, the resulting outputs can be used to establish the robustness of the QNN model.

\balance

\bibliographystyle{IEEEtran}
\bibliography{IEEEabrv,refs.bib}

\end{document}